\title{Essay \LaTeXe{} Template}
\documentclass[letterpaper]{article}
\usepackage[margin=1.00in]{geometry}
\usepackage{mathtools}
\usepackage{amsmath,amsfonts,amssymb,amsthm,hyperref,url}
\usepackage{idxlayout}
\usepackage{makeidx}
\usepackage{sectsty}
\usepackage{bibspacing}
\usepackage{setspace} 
\usepackage[toc,page]{appendix}

\title{\bf\Large{A Quantum Information Geometric Approach to Renormalization} }
\author{\smallskip\large{John B. DeBrota\footnote{\texttt{jdebrota@gmail.com}. This work was done at the Perimeter Institute for Theoretical Physics, 31 Caroline Street North, Waterloo, Ontario N2L 2Y5, Canada.}}\\
    \normalsize{University of Massachusetts Boston} \\ 
\textit{\normalsize{100 Morrissey Boulevard, Boston, MA 02125}}\\
}

\date{\normalsize{\today}}

\newcommand{\brx}[1]{\left(#1\right)}

\begin{document}
\maketitle
\begin{abstract}
This essay constitutes a review of the information geometric approach to renormalization developed in the recent works of B\'eny and Osborne as well as a detailed work-through of some of their contents. A noncommutative generalization of information geometry allows one to treat quantum state distinguishability in geometric terms with an intuitive empirical interpretation, allowing for an information theoretic prescription of renormalization which incorporates both the condensed matter and quantum field theoretic approaches.
\end{abstract}
\section{Introduction}
C\'edric B\'eny and Tobias Osborne have recently developed an operational approach to effective theory construction and renormalization based on quantum information geometry\footnote{For a comprehensive review of quantum information geometry, see \cite{qig}. For a different quantum information geometry based approach to renormalization, following the works of \cite{jaynes93} and \cite{mitchell67}, see \cite{kostecki16}.}.
In short, this approach is as follows. A Riemannian information metric on the manifold of quantum states can be derived as the second order Taylor expansion of the quantum relative entropy. Using it, given a
channel (completely positive trace-preserving map) representing one's experimental limitations, we obtain a measure of one's reduced ability to experimentally distinguish our initial hypothesis from a nearby state. Idealizing one's information gathering by setting a relevance cutoff, beyond which he cannot distinguish
a direction in state space, allows us to foliate the space into equivalence classes of experimentally indistinguishable states. This provides an effective solution to the inverse problem of generally not being able to identify a unique quantum channel preimage. In addition to aiding in the conceptual development of effective theories, this approach leads to an operational information theoretic interpretation of renormalization. The framework just outlined and its connection to renormalization will be elaborated in this text.

The goal of this essay is to compile and, in many cases, clarify or elaborate on the methods and results laid out in \cite{longbeny} and \cite{shortbeny} (to a much smaller degree we also include \cite{oldbeny} in this list). This paper is meant to be a review accessible to a wide population of theoretical physicists and, as such, the author's contribution has been to thoroughly work through many parts of these papers and clarify and expand where it was deemed appropriate\footnote{Some material described in this essay originates from personal correspondence with C\'edric B\'eny; this information has recently appeared in a new article \cite{newbeny}.}.
In \hyperref[etr]{Section 2} we will discuss effective theories and renormalization a bit more. Then, in \hyperref[qcrerm]{Section 3}, we will introduce the elementary mathematical tools of quantum information geometry, required for a further discussion. In \hyperref[boar]{Section 4} we will present the basic concepts of the B\'eny--Osborne approach to renormalization. In \hyperref[cp]{Section 5} we begin to see some features of renormalization in a classical single particle toy model. In
\hyperref[cf]{Section 6}, this analogy is extended to a classical field scenario. In \hyperref[psqt]{Section 7}, we review a few aspects of phase space quantum theory in preparation for the subsequent sections. In \hyperref[cf2]{Section 8}, we revisit the classical field model using in the language presented in Section 7. In \hyperref[details]{Section 9}, we discus technical details of applying Section 7 to a quantum situation, and,
in \hyperref[qp]{Section 10}, we briefly do so for a single quantum particle. In \hyperref[qf]{Section 11}, we address applying the whole formalism to quantum scalar fields. In \hyperref[renorm]{Section 12}, we fully revisit the topic of renormalization with the additional perspectives gained in the previous sections. Finally, we wrap things up in \hyperref[disc]{Section 13} and look towards the future.

\section{Effective Theories and Renormalization}\label{etr}
Physics is essentially empirical. As scientists, we want to build models for particular situations that allow us to co\"ordinate our expectations for unperformed experiments with our knowledge of those that were performed. It is thus essential that we have a systematic way to make good predictions despite experimental limitations. When choosing a quantum state to ascribe to a system, an experimenter must be
mindful of the association between theory and experimental context he has made and his consequent operational ability to distinguish between quantum state assignments. In general, the lack of a well defined procedure to accomplish this could lead to ambiguities in effective theory construction\footnote{Popular accounts often seem to suggest that physics is ``nearly complete''; that all remaining developments in physics will amount to minor tweaks to the currently adopted formalisms and that the work of the next generations will mainly be to work out various emergent properties. However, this perspective idealizes the current state of the art. If we refrain from this attitude, we can consider all existing theories as effective approximations of yet unknown future theories with a wider experimental applicability. The construction of a really good effective theory for one aspect of our experience is hardly something to be ashamed of---Newton's Law of Universal Gravitation is relevant only insofar as we can think of our system as being comprised of several massive bodies (analogy from \cite{fuchs}). With such a perspective we won't fixate on whether we can find a grand unified theory or feel compelled to disparage other disciplines as derivative \cite{xkcd}.}.

What do we mean by an effective theory? One reading of the modifier ``effective'' suggests that an effective theory should be thought of as a remedy that we reluctantly adopt in the absence of the fundamental theory. This is a pessimistic perspective as it reduces an effective theory to ``just'' a step along the path towards a supposed Platonic fixed point. Another reading of ``effective'' designates effective theories as ``theories that work (within a prescribed range of parameters)''. This reading aligns with science's empirical roots: for an experimenter, every theory may as well be an effective theory as the main point of an experiment is to see how well a theory works within the context of experiments he is actually able to perform. In this context, we may find ourselves wanting to construct a theory for large things out of a theory we already have for small things. Or we might want to explore in what ways our best theory for small things could be interpreted as an effective theory derivable from theories of even smaller things. Perhaps we are interested more generally in the way theories at different scales must mesh together. In all of these cases, we must study effective theories themselves in order to understand how and why effective laws can emerge from consistency conditions. 

The concept of the renormalization group\footnote{Groups are not actually involved; ``semigroup'' would be the proper terminology. In fact, all three words ``the'', ``renormalization'', and ``group'' are improper choices which remain in use for historical reasons \cite{blake}.} appeared in the 70s and borrowed the term ``renormalization'' from earlier methods that had appeared in quantum field theory (QFT) \cite{stu&pete53,gell&low54}. What is meant by the renormalization group, while similar to the original ideas, is motivationally distinct from them; the renormalization group concept arose in condensed matter, not particle physics, and found its first applications in the theory of phase transitions. In particle physics, renormalization refers to determining how bare\footnote{``Bare'' coupling constants are the values that go into a Lagrangian. For each bare coupling there is a ``physical'' coupling constant which is the value that a theory predicts will be measured (or trivially related to it).} coupling constants must change as the frequency (distance) cutoff is taken to infinity (zero) such that the physical coupling constants are unchanged\footnote{If this limit exists using finitely many bare coupling constants then the theory is ``renormalizable.''}. In condensed matter, we fix the bare coupling constants and a shortest finite length (so we never take the continuum limit) and compute how the physical coupling constants change as we change the distance (or momentum) at which we measure them.
The condensed matter procedure is due to Wilson \cite{wilson} and is more in line with the preferred conception of an effective theory mentioned above, especially in the context of Kadanoff block-decimation \cite{kadanoff}. Although the goals are different in the two scenarios, Wilson's perspective unites them as activities one may undertake within the same arena. Renormalization methods have also found wide application extending beyond strictly physics, for example, in evolutionary dynamics \cite{blake} and various subfields of pure mathematics (such as geometry, combinatorics, and number theory) \cite{confluences}. For these reasons, and because our chief goal is to achieve an even
more unified perspective, we will refer to all contexts involving renormalization-like procedures as
``renormalization'' without qualification in the remainder of this text. 

The perspective we are striving for is this: renormalization itself, in all of its guises, is a method of effective theory construction. Since, in the condensed matter situation, renormalization clearly works by ignoring or throwing out some information about the system, one hopes that a broader information theoretic framework can be developed which treats renormalization as it is practiced both in particle physics and condensed matter in the same way. The goal is similar in spirit to what E.T.~Jaynes \cite{jaynes79,jaynes93} did in the foundations of statistical mechanics. An information theoretic foundation for renormalization would broaden the scope of present understanding and undoubtedly pave the way for future progress.

\section{Quantum Channels, Relative Entropy, and Riemannian Metric on States}\label{qcrerm}

A \textit{density matrix}, often denoted $\rho$, is a positive semi-definite trace-one matrix which encodes an agent's probabilities for outcomes $i$ from an index set $I$ through the \textit{Born rule}
\begin{equation}
    p(i)=\text{Tr}(\rho A_i),
\end{equation}
where $\{A_i\}$ is a set of positive semi-definite matrices which sum to the identity, known collectively as a \textit{positive operator valued measure} (POVM), which correspond to the possible outcomes of a given experiment. We will often refer to a density matrix as a ``quantum state'' or just a ``state''. A \textit{quantum channel} is a completely positive trace-preserving map (CPTP); this is the most general transformation which sends a density matrix to another density matrix. 

B\'eny and Osborne phrase their formalism in terms of a communication channel between two physicists, Alice and Bob. Alice has associated a state to a preparation procedure which she repeatedly performs and sends the associated system to Bob. Bob performs quantum state tomography on the series of incoming signals he receives from Alice, eventually associating the state $\sigma$ with this preparation (the preparation for him is waiting with his instruments ready for an incoming signal from Alice). Suppose that Bob has encoded his expectation for
information losses incurred by this communication in the quantum channel $E$. By means of his state $\sigma$ and his channel $E$, Bob postulates a set of states $\rho$ which would be compatible with his current assignment under his information degradation model, that is, for any element $\rho$ of the family,
$E(\rho)=\sigma$. In this sense, Bob's current state $\sigma$ is a state within an effective theory for the hypothetical model construction situation within which one might assign $\rho$ to a preparation. Although he cannot know the state Alice associates with the preparation procedure, he knows the system she prepares has not been subjected to the losses inherent in the transmission procedure and so, within the confines of his postulates, Bob loosely thinks of the
set of compatible states as the result of ``inverting'' the
quantum channel $E$ to produce a set from which Alice chose her quantum state assignment (of course, the state Alice actually assigned is impossible for Bob to determine). As there is no way to exactly invert a general CPTP map, the question B\'eny and Osborne pose is: how does Bob proceed in view of this inverse inductive inference problem?

We prefer to speak in terms of an agent and a system he has access to. The preceding paragraph requires minor tweaks to pose the question in this way. Consider an agent and a system. Insofar as the agent hopes to perform repeated tests ``on'' the system, he must identify some aspect of it or some sequence of actions he can take which he feels he can treat as an exchangeable\footnote{``Exchangeable'' is subjective Bayesian terminology. Quoting Frank Lad's book \cite{lad}, ``[Y]ou regard a sequence of $N$ quantities exchangeably if your probabilities for observing any two sequences of observation values are equal whenever the components of one
sequence of observations is a mere \textit{permutation} of the components of another.'' From the Bayesian perspective of E.T.~Jaynes, one would speak of a ``testable'' \cite{jaynes68}  preparation procedure instead.} preparation procedure. Suppose that with knowledge of his equipment, his
calibration processes, the associations between experimental context and symbolics he has chosen, and the frequencies of outcomes of his experiments, our agent has, at the conclusion of his tests, assigned the state $\sigma$ to this preparation procedure for the system. The agent additionally postulates a quantum channel $E$ representing his model for how the quantum states constructed according to this procedure are related to (are obtainable from) the quantum states that would be 
constructed according to the unknown but refined procedures and experimental data. Consider now the case where the agent is interested in what ways he could or perhaps should refine his state assignment in case his experimental capabilities improve in some way. As in the previous paragraph, the agent postulates a set of states $\rho$ compatible under $E$. Note that there need not be any meaningful connection between the model construction procedures or between the experimental technologies in the current and imagined
scenarios; he merely postulates that the \textit{state assignments} in each case are compatible under this channel. The refining process, may now be loosely treated as ``inverting'' $E$ to obtain candidates for a refined state $\rho$. As mentioned before, we cannot invert $E$ in general so the agent needs a systematic and sensible way to continue his efforts to improve his state construction. 

In face of the impossibility to provide an exact solution of the inverse problem, any approximate solution will crucially depend on the criteria applied to select what is ``the best'' approximate solution. The central issue is distinguishability of states. How do we quantify our ability to distinguish two states? In principle, it would be nice to have a quantity like a distance that takes two density operators and outputs a number representing how ``far" they are from each other. The two obvious requirements for a notion of distinguishability as a distance are that it is a function of two states that returns zero when both states are the same and is never negative (because no state can be less
distinguishable from another state than it is from itself). One reasonable choice is the Umegaki distance function \cite{umegaki61,umegaki62}, which is the negative of the quantity called the quantum relative entropy
\begin{equation}
    D(\rho,\sigma):=-S(\rho,\sigma)=\text{Tr}(\rho\log\rho-\rho\log\sigma).
\end{equation}
The idea of negative entropy as an ``amount of information'' is due to Wiener \cite{wiener}, and, from this perspective, negative relative entropy is naturally interpreted as a nonsymmetric ``information distance'' \cite{rick2} as it satisfies the two desired conditions for a distinguishability distance enumerated above. As the relative entropy increases, the distinguishability decreases until the states are so ``close together'' that they are no longer noticeably distinct.
It should be noted that the
sign convention we use for relative entropy is the opposite of what is found in many other sources (but it is in agreement with the convention used in \cite{bb79}). The sign of the relative entropy, as used here, preserves the concept of entropy being nondecreasing under a generalized evolution process. For our purposes the choice of the Umegaki distance is mostly arbitrary, and, in fact, the following holds for a
large class of relative entropies \cite{lr99}. One motivation, however, is that in the
commutative case, i.e.,\ when the set of permitted density matrices are simultaneously diagonalizable so that their diagonal elements form probability vectors, the Umegaki distance reduces to the well known Kullback-Leibler divergence, $D_{KL}(p,q)=\sum_i p_i \log{(p_i/q_i)}$, which will allow us to directly compare quantum and classical situations with minimal difficulty in our examples later on. 

Since our quantum channels are meant to include the effects of noise, the distinguishability of two states should be nonincreasing under the action of the channel, 
\begin{equation}\label{monotone}
    D(\rho,\sigma)\geq D(E(\rho), E(\sigma)).
\end{equation}
For a quantum channel $E$ with nontrivial kernel, two states $\rho$ and
$\sigma$ are called \textit{equivalent} if $E(\rho)=E(\sigma)$. In practice, exact equivalence would be vanishingly rare so we loosen the condition: two states $\rho$ and $\sigma$ will be called \textit{approximately equivalent} if $D(E(\rho), E(\sigma))\leq \epsilon$ for some $\epsilon>0$ chosen by the agent based upon his desired degree of distinguishing confidence and experimental capabilities. 

In general, with respect to a given state, nearby states along some paths of the set will contract more than those along others under the action of $E$. Physically this could be because of peculiarities of the measuring apparatus or the lab environment. In general, the preimage of an $\epsilon$ ball will be a pancake shaped set of approximately equivalent states. The inverse problem may be informally solved as follows: decide an amount of contraction to act as a threshold, beyond which we idealize the contraction all the way to the exactly equivalent case, leaving us with lower dimensional sheets that foliate the original manifold into equivalence classes. A set of effective states would then be a curve that intersects each equivalence sheet exactly once and specifies a unique preimage for $E$. However, the relative entropy is difficult to calculate in practice so B\'eny and Osborne take a different approach that is motivated by this flattened pancakes idea, using a Riemannian metric derived from the relative entropy.

In finite dimensions, the set of strictly positive quantum states may be given a natural manifold structure simply by parametrizing the states by open subsets of $\mathbb{R}^n$ where $n$ is the dimension of the Hilbert space. In infinite dimensions, this approach does not work, but it is still possible to obtain manifold structure by using noncommutative Orlicz spaces as local tangent spaces instead of $\mathbb{R}^n$ \cite{jencova}. We informally assume a manifold structure for the remainder of this paper.

Instead of working with arbitrary points on our manifold of states, we assume there is a hypothesis state $\rho$ and the agent is concerned with sorting out the finer details in a small neighborhood of this point. In other words, we want to know how this nonsymmetric information distance behaves infinitesimally. In order for $\rho+\epsilon X$ to be a valid point on our manifold, the $X$ must be a traceless Hermitian matrix. Following \cite{longbeny} we call these $X$ \textit{features}. To obtain the infinitesimal behavior, we would like to Taylor expand our distance function. This is a nontrivial task. First we state the result applied to our hypothesis $\rho$ and the nearby state $\rho + \epsilon X$ and then explain the derivation. We get to lowest order in $\epsilon$:
\begin{equation}\label{neighborhood}
D(\rho+\epsilon X, \rho) = \epsilon^2\text{Tr}(X \Omega_\rho^{-1}(X))+\mathcal{O}\left(\epsilon^3\right)
\end{equation}
where
\begin{equation}\label{metric}
    \Omega_\rho^{-1}(Y):=\frac{d}{dt}\biggr\rvert_{t=0}\log(\rho+tY).
\end{equation}
We can think of $\Omega_\rho^{-1}$ as a noncommutative ``division by $\rho$'' operator due to its behavior in the commutative case; if $\rho$ and $Y$ were real functions, $\Omega_\rho^{-1}(Y)=\frac{Y}{\rho+tY}\bigr|_{t=0}=Y/\rho$.

Although at this stage it is not easy to see, $\Omega^{-1}_\rho$ is self-adjoint in the Hilbert--Schmidt inner product,
$(A,B)_{HS}:= \text{Tr}(A^\ast B)$. Further, we note that $\Omega_\rho^{-1}(\rho)=\mathbb{I}$ for all $\rho$, which follows if we diagonalize $\rho$ and observe that each entry gives us $\frac{\lambda}{\lambda+t\lambda}\bigr|_{t=0}=1$. The Taylor expansion about $\{\rho,\rho\}$ takes the form

\begin{equation}\label{taylor}
\begin{split}
D(p&,q)=D(\rho,\rho)+\left(|\!|\Delta p|\!|\Bigr\lvert\!\Bigr\lvert\mathcal{D}_pD(p,q)\Bigr\rvert_{\substack{p=\rho\\q=\rho}}\Bigr\rvert\!\Bigr\rvert+|\!|\Delta q|\!|\Bigr|\!\Bigr|\mathcal{D}_qD(p,q)\Bigr\rvert_{\substack{p=\rho\\q=\rho}}\Bigr|\!\Bigr|\right)\\
&+\frac{1}{2}\left(|\!|\Delta p|\!|^2\Bigr|\!\Bigr|\mathcal{D}_p\mathcal{D}_pD(p,q)\Bigr\rvert_{\substack{p=\rho\\q=\rho}}\Bigr|\!\Bigr|+2|\!|\Delta p|\!||\!|\Delta q|\!|\Bigr|\!\Bigr|\mathcal{D}_p\mathcal{D}_qD(p,q)\Bigr\rvert_{\substack{p=\rho\\q=\rho}}\Bigr|\!\Bigr| + |\!|\Delta q|\!|^2\Bigr|\!\Bigr|\mathcal{D}_q\mathcal{D}_qD(p,q)\Bigr\rvert_{\substack{p=\rho\\q=\rho}}\Bigr|\!\Bigr| \right)\\
&+\mathcal{O}\left(|\!|\Delta\{p,q\}|\!|^3\right)
\end{split}
\end{equation} 
where $\Delta p$ and $\Delta q$ are displacements from $\rho$, $\mathcal{D}_n$ is the Fr\'echet derivative of index $n$, and $|\!|\cdot|\!|$ is the operator norm. If $f$ is Fr\'echet differentiable at an operator $U$, then \cite{bhatia}, for all operators $V$, 
\begin{equation}
\mathcal{D}_Uf(V)=\frac{d}{dt}\biggr\rvert_{t=0}f(U+tV),
\end{equation}
 and so we see that $\Omega^{-1}_\rho(Y)$ is the Fr\'echet derivative of the $\log$ function at $\rho$ in the direction of $Y$.

The first term of the Taylor expansion is $D(\rho,\rho)=0$ because the distance between a point and itself is always zero. We also expect the $\mathcal{O}(\epsilon)$ term to vanish because we're expanding our function about a minimum ($D$ is nonnegative and $D(\rho,\sigma)=0 \text{ iff } \rho=\sigma$). Indeed, taking the first single derivative term in \eqref{taylor}, we obtain
\begin{equation}
\frac{d}{dt}\text{Tr}\left((\rho+tX)\log(\rho+tX)-(\rho+tX)\log\rho\right)\Bigr|_{t=0}=\text{Tr}\left(\rho\Omega^{-1}_\rho(X)\right)=\text{Tr}\left(\Omega^{-1}_\rho(\rho)X\right)=\text{Tr}(X)=0,
\end{equation}
where we used the self-adjointness of $\Omega^{-1}_\rho$ and that $X$ is traceless. The other term similarly vanishes as expected. The first term that is not obviously trivial is the second order term. It is composed of three second partial derivative terms. The first pure second partial derivative term gives us the following: 
\begin{equation}
    \begin{split}
        &\frac{\partial^2}{\partial s\partial t}D\left(\rho +t X+s Y,\rho\right)\Bigr\rvert_{\substack{s=0\\t=0}}=\frac{\partial^2}{\partial s\partial t}\text{Tr}\left((\rho+t X+s Y)\log(\rho +t X+s Y)-(\rho+t X+sY)\log\rho\right)\Bigr\rvert_{\substack{s=0\\t=0}}\\
        &=\frac{\partial}{\partial s}\text{Tr}\left(\rho \frac{\partial}{\partial t}\log(\rho+tX+sY)\Bigr\rvert_{t=0}+X\log(\rho+sY)+sY\frac{\partial}{\partial t}\log(\rho+tX+sY)\Bigr\rvert_{t=0}-X\log\rho\right)\Bigr\rvert_{s=0}\\
        &=\text{Tr}\left(\rho\frac{\partial^2}{\partial s\partial t}\log(\rho+tX+sY)\Bigr\rvert_{\substack{s=0\\t=0}}+X\frac{\partial}{\partial s}\log(\rho+sY)\bigr\rvert_{s=0}+Y\frac{\partial}{\partial t}\log(\rho+tX)\bigr\rvert_{t=0}\right)\\
        &=\text{Tr}\left(\rho\frac{d}{d s}\Omega_{\rho+sY}^{-1}(X)\Bigr\rvert_{s=0}\right)+\text{Tr}(X\Omega_\rho^{-1}(Y))+\text{Tr}(Y\Omega^{-1}_\rho(X)),
    \end{split}
\end{equation}
and the other pure second partial derivative term gives us the same answer with the opposite sign, so it cancels out when we choose a uniform displacement $\epsilon$. The mixed second partial derivative term does not vanish:
\begin{equation}\label{fintaylor}
\begin{split}
&\frac{\partial^2}{\partial s\partial t}D\left(\rho +t X,\rho+sY\right)\Bigr\rvert_{\substack{s=0\\t=0}}=\frac{\partial^2}{\partial s\partial t}\text{Tr}\left((\rho+t X)\log(\rho +t X)-(\rho+t X)\log(\rho+sY)\right)\Bigr\rvert_{\substack{s=0\\t=0}}\\
&=\frac{\partial}{\partial s}\text{Tr}\left(\rho\frac{d}{dt}\log(\rho+tX)\Bigr|_{t=0}+X\log(\rho+tX)-X\log(\rho+sY)\right)\Bigr|_{s=0}=-\text{Tr}(X\Omega_\rho^{-1}(Y)).
\end{split}
\end{equation} 
Plugging this into \eqref{taylor}, idealizing to the situation where we only care about the direction $X$, and fixing $\epsilon$, gives us \eqref{neighborhood} except that the left hand side is $D(\rho+\epsilon X,\rho+\epsilon X)$ instead of $D(\rho+\epsilon X,\rho)$. The reason this happened is manifest in the derivation: the \textit{direction} from our hypothesis point is the aspect of the distinguishability that matters at the infinitesimal scale. In that sense, what we described as the distinguishability between $\rho$ and $\rho+\epsilon X$ is more meaningfully thought of in terms of a magnitude of the ``tangent vector'' $X$. Let's make this claim more explicit. 

Consider all possible paths in the manifold $\mathcal{M}$ originating at $\rho$. The tangent space $T_\rho\mathcal{M}$ is the set of all tangents to these curves at $\rho$ abstractly represented by directional derivative operators. The operators $X$ from before may be meaningfully be identified with elements of the tangent space $T_\rho\mathcal{M}$ in a natural way. First, fix a basis such that an arbitrary vector is written $U=\sum_iU_i\frac{\partial}{\partial x_i}$ where $\frac{\partial}{\partial x_i}$ represents all possible co\"ordinate derivatives and $U_i$ are the components in this basis. $\rho+\epsilon X$ is a point on the manifold so there is a path $s(t)$ originating at $\rho$ which passes through $\rho+\epsilon X$ at time $t=\epsilon$, i.e.,\ $s(\epsilon)=\rho+\epsilon X$. Consider an arbitrary differentiable real-valued function $f$ which acts on the manifold and Taylor expand the evaluation of $f$ along $s(t)$ about $t=0$:
\begin{equation}
f(s(t))=f(\rho)+t\sum_{i}s'_{i}(0)\frac{\partial f(\sigma)}{\partial x_{i}}\biggr|_{\sigma=\rho}+\mathcal{O}\left(t^2\right).
\end{equation} 
At $t=\epsilon$ we then have 
\begin{equation}
f(s(\epsilon))=f(\rho+\epsilon X)=f(\rho)+\epsilon\sum_{i}s'_{i}(0)\frac{\partial f(\sigma)}{\partial x_{i}}\biggr|_{\sigma=\rho}+\mathcal{O}\left(\epsilon^2\right).
\end{equation}
Thus, to first order in $\epsilon$,  
\begin{equation}
X=\sum_{i}s'_{i}(0)\frac{\partial}{\partial x_{i}}\in T_\rho\mathcal{M}.
\end{equation}

Thinking of the $X$ operators as tangent vectors allows us to extract more content from the expansion of the distance function. Ingarden \textit{et al}.\ \cite{ijkk} showed that this expansion defines a norm from which we can construct a Riemannian metric on our manifold  (for derivations of different Riemannian metrics from a broad class of quantum relative entropies, see \cite{lr99}). The Bogolyubov--Kubo--Mori (BKM) \cite{bogo,kubo,mori} inner product is defined as the negative of the result in \eqref{fintaylor},
\begin{equation}\label{bkm}
\langle A, B \rangle_\rho := \text{Tr}(A^\ast \Omega_\rho^{-1}(B)).
\end{equation}

Armed with a metric, we can see exactly how much a direction shrinks under $E$ and thus formulate a measure of the importance of a feature. We define the \textit{relevance} of a feature $X$ as the ratio:
\begin{equation} \label{eq:relevanceratio}
 \eta_\rho(X) := \frac{\langle E(X),E(X) \rangle_{E(\rho)}}{\langle X,X \rangle_\rho}.
 \end{equation}
A relevance close to $1$ indicates that states nearby the hypothesis in the direction of that feature remain easy to distinguish after the application of $E$. 

A simple toy example that is nonetheless important to keep in mind is the partial trace channel. This is important because partial tracing is the prototype of what we do when we throw out information. If a system consists of two qbits and an agent only has access to one of them, then the quantum channel he should use to model his inadequacy is the partial trace over the subsystem he cannot measure. Formally, if the unknown state the agent associates to the system is $\rho_{AB}$ and he can only
access subsystem A, then he models this limitation via $E(\rho_{AB})=\text{Tr}_B(\rho_{AB})=:\rho_A$. If the agent's hypothesis state is the completely mixed state, $\rho_{AB}=\mathbb{I}\otimes\mathbb{I}/4$, then, since $[\rho_{AB},X]=0$ for any feature $X$, $\Omega^{-1}_{\rho_{AB}}(X)$ becomes $X\rho_{AB}^{-1}=4X$. It is then easy to calculate \eqref{eq:relevanceratio}: he will conclude that any feature of the form $P\otimes\mathbb{I}$ has relevance 1 and any feature of the form
$P \otimes Q$ with Tr$(P)=\text{Tr}(Q)=0$ has relevance 0.

In principle, the agent may now calculate the $n$ most relevant features at a particular $\rho$ given a quantum channel $E$. If he orders them in decreasing relevance then after a certain cutoff, he may say that his experiments are simply not sophisticated enough to distinguish a state from one very nearby in a direction of a feature beyond the cutoff. All features beyond the cutoff are called \textit{irrelevant} and those before the cutoff are \textit{relevant}. Now we have an exact equivalence
relation; if the difference between two states is irrelevant at $\rho$ (to first order) then the states are in the same equivalence class. How do we calculate the $n$ most relevant features? We will see below that this problem, while still difficult, is made more tractable after we introduce the adjoint.

Corresponding to a feature $X$, define the \textit{observable} $A=\Omega_\rho^{-1}(X)$. Since $\text{Tr}\left((\cdot)^*\Omega_\rho^{-1}(\cdot)\right)$ is a Riemannian metric, the operator $A$ is a covector. We could similarly write $X$ in terms of $A$ as $X=\Omega_\rho(A)$ although we do not yet have an explicit expression for $\Omega_\rho$. Consider an arbitrary self-adjoint operator $H$ and form $e^{-H}$. Suppose we perturb $H$ by $\epsilon A$ and Taylor expand about $-H$:
\begin{equation}
    e^{-H+\epsilon A}=e^{-H}+\epsilon \frac{d}{dt}\biggr|_{t=0} e^{-H+tA}+\mathcal{O}\left(\epsilon^2\right).
\end{equation}
Take the log of both sides and Taylor expand about $e^{-H}$,
\begin{equation}
    \begin{split}
        -H+\epsilon A&=\log\left(e^{-H}+\epsilon \frac{d}{dt}\biggr|_{t=0} e^{-H+tA}+\mathcal{O}\left(\epsilon^2\right)\right)\\
                     &=\log(e^{-H})+\epsilon\frac{d}{ds}\biggr|_{s=0}\log\left(e^{-H}+s\frac{d}{dt}\biggr|_{t=0} e^{-H+tA}\right)+\mathcal{O}\left(\epsilon^2\right)\\
                     &=-H+\epsilon \Omega^{-1}_{e^{-H}}\left(\frac{d}{dt}\biggr|_{t=0} e^{-H+tA}\right)+\mathcal{O}\left(\epsilon^2\right).
\end{split}
\end{equation}
Matching terms to first order in $\epsilon$ implies that $\Omega_{e^{-H}}(A)=\frac{d}{dt}\bigr|_{t=0} e^{-H+tA}$, the Fr\'echet derivative of the exponential function at $-H$ in the direction $A$. If we introduce the \textit{partition function} $Z:=\text{Tr}(e^{-H})$, then $\rho=e^{-H}/Z$ is a valid quantum state, called a \textit{thermal state}, and $H$ may be interpreted as a Hamiltonian. Given a Hamiltonian, there is a corresponding thermal state. Conversely, given $\rho$, there is a
Hamiltonian for which $\rho$ is the thermal state. Observe the following relation:
\begin{equation}\label{swapz}
    \Omega^{-1}_\rho(X)=\frac{d}{dt}\biggr|_{t=0}\log\left(\frac{e^{-H}}{Z}+tX\right)=\frac{d}{dt}\biggr|_{t=0}\left(\log(1/Z)+\log\left(e^{-H}+tZX\right)\right)=Z\Omega^{-1}_{e^{-H}}(X),
\end{equation}
and so
\begin{equation}\label{omegap}
    \Omega_\rho(A)=\frac{1}{Z}\frac{d}{dt}\biggr|_{t=0}e^{-H+tA}.
\end{equation}
Also note that $Z$ is invariant to first order in $\epsilon$ under perturbations by an observable:
\begin{equation}
    \begin{split}
        \text{Tr}\left(e^{-H+\epsilon A}\right)&=\text{Tr}\left(e^{-H}+\epsilon\Omega_{e^{-H}}(A)+\mathcal{O}\left(\epsilon^2\right)\right)\\
                                               &=Z+\epsilon\text{Tr}\left(\Omega_{e^{-H}}\left(\Omega^{-1}_\rho(X)\right)\right)+\mathcal{O}\left(\epsilon^2\right)\\
                                               &=Z+\epsilon Z \text{Tr}(X)+\mathcal{O}\left(\epsilon^2\right)\\
                                               &=Z+\mathcal{O}\left(\epsilon^2\right)
\end{split}
\end{equation}
where we used relation \eqref{swapz} in the penultimate step. Putting it all together,
\begin{equation}
    \frac{1}{Z}e^{-H+\epsilon A}=\frac{1}{Z}e^{-H}+\frac{\epsilon}{Z}\Omega_{e^{-H}}(A)+\mathcal{O}\left(\epsilon^2\right)=\rho+\epsilon \Omega_\rho(A)+\mathcal{O}\left(\epsilon^2\right)=\rho+\epsilon X+\mathcal{O}\left(\epsilon^2\right),
\end{equation}
i.e.,\ we may think of perturbations to a state by a feature $X$ as equivalent to perturbations to the Hamiltonian defining the corresponding thermal state by the related observable $A$.

\section{B\'eny--Osborne Approach to Renormalization}\label{boar}
We now define \textit{the adjoint $\mathcal{R}_\rho$ of channel $E$} under the BKM inner product \eqref{bkm}. This adjoint is not to be confused with the Hilbert--Schmidt adjoint, $E^\ast$. $\mathcal{R}_\rho$ is defined by the following relation:
\begin{equation}
\langle\mathcal{R}_\rho(Y),X\rangle_\rho=\langle Y,E(X) \rangle_{E(\rho)}, 
\end{equation}
where $Y$ is a feature at $E(\rho)$ and $X$ is a feature at $\rho$. This notion of adjointness relates the inner products at $\rho$ and at $E(\rho)$. Let's see if we can motivate this a bit more transparently. Using the definition of the metric, we can explicitly calculate it to be 
\begin{equation} \label{eq:adjrho}
\mathcal{R}_\rho=\Omega_\rho E^\ast \Omega_{E(\rho)}^{-1}. 
\end{equation}
If we look at the action of $\mathcal{R}^\ast_\rho=\Omega_{E(\rho)}^{-1} E \Omega_\rho$ on an observable associated with $\rho$, we see that an observable is pulled back to a feature at $\rho$ under the action of $\Omega_\rho$, mapped to the corresponding tangent space at the point $E(\rho)$ under $E$ (we may also think of $E$ as a map from a space of equilibrium states of the same temperature to another point on the same manifold thus defining an ``$E$-flow"), and finally pushed forward to a vector in the space of observables associated with $E(\rho)$. So $\mathcal{R}^\ast_\rho$ accomplishes on observables what $E$ accomplishes on features.

Knowledge of the adjoint makes it easier to compute the relevance because we can write the inner product at $E(\rho)$ in terms of the inner product at $\rho$,  
\begin{equation}\label{shifttorho}
\langle E(X),E(Y) \rangle_{E(\rho)}=\langle X, \mathcal{R}_\rho (E(Y))\rangle_\rho.
\end{equation}
This makes things nicer because we've pushed the agent's limitations into an operator so we can now work with just the inner product at $\rho$. In fact, the problem is immediately seen to be a generalized eigenvalue relation. In \eqref{eq:relevanceratio}, multiply through by the denominator on the right hand side and substitute in \eqref{shifttorho} for the inner product at $E(\rho)$ in the case where $Y=X$. We see that finding the $n$ most relevant features is a matter of solving the following for $X_n$:
\begin{equation}
\mathcal{R}_\rho E(X_n)=\eta_n X_n.
\end{equation}
For observables the corresponding eigenrelevance relation can similarly be derived to be  
\begin{equation} \label{eq:dualpicture}
    E^\ast \mathcal{R}_\rho^\ast (A_n) = \eta_n A_n.
\end{equation} 
The notion of equivalence carries over as one might expect to this dual picture. An observable is relevant if it belongs to the span of the first $n$ eigenrelevance observables. Two states, $\rho_1$ and $\rho_2$ are equivalent to first order if they yield the same expectation values for all of the relevant observables. This is a nice definition because comparing expectation values is one of the few things one can do to try to distinguish quantum systems.

As we will see, this formalism allows for an information theoretic characterization of renormalization which includes both the condensed matter and QFT perspectives. Specifically, the applicability and success of the renormalization procedure in all of its guises is postulated to be a consequence of the freedom we have to choose a representative within the equivalence class of indistinguishable states generated by a given hypothesis. In condensed matter, traditional renormalization flow towards simpler Hamiltonians which preserve some desired features of a system, e.g., long distance correlations, can be understood as the process of choosing a new Hamiltonian such that the corresponding thermal state remains within the equivalence class of the original state for a chosen relevance cutoff. In QFT, renormalization traditionally refers to the parameter flow determined by the condition that the predictions of the theory are independent of the regulator. This prescription is seen to be consistent with parameter flow as a function of the regulator being determined by the requirement that renormalized states remain within the original equivalence class. Furthermore, divergences are seen to result from the presence of infinitely many irrelevant features, so the inclusion of a regulator, which amounts to removing the irrelevant features, obtains a satisfying information theoretic justification as well.

\section{Classical Particle}\label{cp}

We start with a classical single particle model system to illustrate the general idea of this approach before we build the machinery necessary to attempt a more sophisticated system. We will identify some aspects of renormalization in the context of this model as well.

The setting is an inference problem for some agent. He wishes to experimentally infer a probability distribution for some system that has been subjected to noise modeled by a known stochastic map $E$. 

Although in a general theory we'd use density matrices or phase space equivalents, in a classical single particle model, we want a commuting representation so states in this model are probability distributions $p(x)$ over $\mathbb{R}$ and $E$ is convolution (denoted $\star$) by a Gaussian normal distribution $N$:
\begin{equation} \label{eq:convolution}
E(p)(x):= \brx{N \star p}(x)=\frac{1}{\sqrt{2\pi}\sigma}\int_{-\infty}^{\infty} p\brx{y}e^{-\frac{1}{2\sigma^2}(x-y)^2}dy.
\end{equation}

Convolution with a Gaussian for some probability distribution will look like smoothing of corners and broadening of peaks. It is chosen because it represents a reasonable averaged noise process and also because Gaussians are easy to work with, but in principle any stochastic map may be chosen. The parameter $\sigma$ is the width (variance) of the distribution and is one way to represent the uncertainty the agent assigns to his measurements of the variable $x$. 

For our formalism the agent needs an initial hypothesis which we also take to be a normal distribution. If this agent has no information other than some idea of an average value for the particle's position then, assuming he wants to construct a normalized probability distribution, he chooses the entropy-maximizing normal distribution \cite{jaynes} as his prior (centered at $0$ without loss of generality),
\begin{equation}
p(x)=\frac{1}{\sqrt{2\pi}\tau}e^{-\frac{x^2}{2\tau^2}}.
\end{equation}
We want to compute the eigenrelevance directions about this state. In order to do this we need to solve the eigenvalue equation $E^\ast \mathcal{R}_\rho^\ast (A_n) = \eta_n A_n$. First we determine $E^\ast$, the adjoint of $E$ defined by $\brx{X,E(Y)}=\brx{E^\ast(X),Y}$ where we are working in the simple $L^2$ space over $\mathbb{R}$ because probability distributions are normalized real-valued functions. We see that $E^\ast$ functionally acts the same way as $E$:
\begin{equation}
\begin{split}
\brx{q,E(p)} & = \frac{1}{\sqrt{2\pi}\sigma}\int q(x) \int p(y)e^{-\frac{1}{2\sigma^2}(x-y)^2}dy dx \\
& = \frac{1}{\sqrt{2\pi}\sigma}\int \int p(x') q(y') e^{-\frac{1}{2\sigma^2}(y'-x')^2}dx dy \\
& = \frac{1}{\sqrt{2\pi}\sigma}\int p(x) \int q(y) e^{-\frac{1}{2\sigma^2}(x-y)^2}dx dy \\
& = \brx{E^\ast(q),p}.
\end{split}
\end{equation}
Where in the second line we made the variable substitutions $ x \to y'$ and $ y \to x'$ and in the third line we sent $x' \to x$ and $y' \to y$ and took advantage of the fact that $(y-x)^2=(x-y)^2$. 

As probability distributions are real-valued functions which commute, operators $\Omega_{E(p)}^{-1}$ and $\Omega_p$ in this case are simply division by $E(p)$ and multiplication by $p$ respectively. Thus $\mathcal{R}_p^\ast(A_n)=\frac{E(A_n p)}{E(p)}$ and 
\begin{equation}
 E^\ast\mathcal{R}_p^\ast(A_n)=E^\ast\brx{\frac{E(A_n p)}{E(p)}}.
\end{equation}
The convolution of a Gaussian with a Gaussian, $E(p)$, is another Gaussian with scaled variance:
\begin{equation}
E(p)(x)=\frac{1}{\sqrt{2\pi(\sigma^2+\tau^2)}}e^{-\frac{x^2}{2(\sigma^2+\tau^2)}}.
\end{equation}
Thus we have 
\begin{equation}
\begin{split}
(E^\ast\mathcal{R}_p^\ast(A_n))(x) & = E^\ast\brx{\sqrt{2\pi\brx{\sigma^2+\tau^2}}e^{\frac{x}{2(\sigma^2+\tau^2)}}\int \frac{A_n(y)}{\sqrt{2\pi}\tau}e^{\frac{-y^2}{2\tau^2}}e^{\frac{-(x-y)^2}{2\sigma^2}}dy}\\
& = E^\ast\brx{\frac{\sqrt{\sigma^2+\tau^2}}{\tau}\int A_n(y)\text{Exp}\brx{-\frac{\brx{y \sigma^2+(y-x)\tau^2}^2}{2\sigma^2\tau^2(\sigma^2+\tau^2)}}dy}\\
& = \frac{\sqrt{\sigma^2+\tau^2}}{\sqrt{2\pi}\sigma\tau}\int\brx{\int A_n(y)\text{Exp}\brx{-\frac{\brx{y \sigma^2+(y-z)\tau^2}^2}{2\sigma^2\tau^2(\sigma^2+\tau^2)}}dy}e^{-\frac{(x-z)^2}{2\sigma^2}}dz.
\end{split}
\end{equation}
To get something sensible from this, it's better to take a step back to the second line and remember the $E^\ast$ operator is a convolution. Thanks to this particular form, we can apply the convolution theorem, 
\begin{equation}
f\star g=\mathcal{F}^{-1}\brx{\mathcal{F}(f)\mathcal{F}(g)},
\end{equation}
where $\mathcal{F}$ is the Fourier transform operator. Especially since we are working with Gaussians, this calculation becomes easy because  $\mathcal{F}\brx{\frac{1}{\sqrt{2\pi}\tau}e^{-\frac{x^2}{2\tau^2}}}= e^{\frac{-k^2 \tau^2}{2}}$. So after inverse Fourier transforming the product, we get our eigenvalue relation:

\begin{equation}
    (E^\ast\mathcal{R}_p^\ast(A_n))(x)= \frac{\sigma^2+\tau^2}{\sigma \tau\sqrt{2\pi(\sigma^2+2\tau^2)}}\int A_n(y)\text{Exp}\brx{-\frac{\brx{y\sigma^2+(y-x)\tau^2}^2}{2\sigma^2\tau^2(\sigma^2+2\tau^2)}}dy=\eta_n A_n(x).
\end{equation}
To get a cleaner relation and instructive eigenvalues we define $\alpha=(\sigma^2+\tau^2)/\tau^2$ and obtain
\begin{equation} \label{eq:toyeval}
    (E^\ast\mathcal{R}_p^\ast(A))(x)=\frac{\alpha}{\sqrt{2\pi(\alpha^2-1)}\tau}\int A(y) e^{-\frac{(x-\alpha y)^2}{2\tau^2(\alpha^2-1)}}dy=\eta_n A(x). 
\end{equation}

The situation looks a bit less painful, but it looks like we're still faced with solving a rather difficult eigenvalue relation---for which $A(y)$ does expression (\ref{eq:toyeval}) give us something proportional to $A(y)$? Certainly for a general situation (a non-Gaussian state subject to a non-Gaussian channel) this would be bad. In this case, however, since the Hermite polynomials can be defined in terms of derivatives of Gaussians, we can see that the eigenvectors of (\ref{eq:toyeval}) are simply properly scaled Hermite polynomials,
\begin{equation}
A_n(x):=\frac{1}{\sqrt{n!}}H_n(x/\tau)=(-\tau)^n\frac{1}{\sqrt{n!}}e^\frac{x^2}{2\tau^2}\frac{d^n}{dx^n}e^{-\frac{x^2}{2\tau^2}},
\end{equation}
and the eigenvalues are
\begin{equation}
\eta_n=1/\alpha^n.
\end{equation}
We may prove this with the generating function of the Hermite polynomials, 
\begin{equation}
f_t(x)=\sum_{n=0}^\infty A_n(x)t^n/n! = e^{xt/\tau-t^2/2},
\end{equation}
which satisfies $E^\ast\mathcal{R}_p^\ast(f_t)=f_{t/\alpha}$. The terms of the Taylor series in $t$ of $f_{t/\alpha}$ reveal the desired eigenvector and eigenvalue relation. 

It is worth mentioning that since the observables are Hermite polynomials, the features turn out to be proportional to the familiar simple harmonic oscillator energy eigenstates. Thinking in these terms may provide an intuition that is useful in field theoretic settings.

The cutoff after which one should deem states to be irrelevant is determined by experimental constraints such as the size and power of a collider or the number of experiments one has time to run. Even in this simple model we will be able to see elements of the renormalization procedure seen from perspectives analogous to those in statistical physics and QFT. First we will consider how this situation may work from the statistical physics picture.

We may, somewhat improperly since there is no mention of a phase space in this example, treat the exponential part of our Gaussian hypothesis state as a ``Hamiltonian", $H=x^2/2\tau^2$, so that the Gaussian hypothesis can be thought of as the thermal state. Assume the agent has some detailed Hamiltonian $H_0=x^2/2\tau^2+\lambda x^4/\tau^4$ (assume for now that $\lambda>0$) near the Gaussian state for the short distance physics of his system. Depending on his experimental capabilities and aims,
however, he may prefer to use a simpler effective theory with a larger minimum length scale of applicability. The thermal state that the agent associates with $H_0$ is simply $p_0=e^{-H_0}$. The renormalization problem for the agent can now be solved by thinking in terms of the thermal state of the Hamiltonian rather than in terms of the Hamiltonian itself. Say, for example, that the agent determines the relevant observables to be those with $n\leq2$. With this knowledge, he may use this
opportunity to find a simpler Hamiltonian $H_1=x^2/\tau_1^2$ associated to a state $p_1=e^{-H_1}$ which is equivalent to $p_0$. $A_1$ is an odd function so its expectation value is zero in both cases. $A_2=-\frac{1}{\sqrt{2}}+\frac{x^2}{\sqrt{2}\tau^2}$ so we require that the second moment (expectation value of $x^2$) is the same for $p_1$ and $p_2$ (the constant term in $A_2$ doesn't matter since both $p_1$ and $p_2$ will be normalized). The variance of $p_0$ is, to first order in $\lambda$, $(1-12\lambda)\tau^2$. So we may choose $\tau_1$ to be the square root (again to first order in $\lambda$), $\tau_1=(1-6\lambda)\tau$ so that $p_1$ is equivalent to $p_0$. The flow from $H_0$ to $H_1$ may be thought of as one step in the renormalization process as understood in statistical physics.

We can also see some aspects reminiscent of the QFT renormalization procedure in this toy model. Instead of looking for a simplified Hamiltonian, in QFT we are often trying to get a more widely applicable effective theory. Say for instance that the agent starts with a Gaussian hypothesis with Hamiltonian $H_0=x^2/2\tau_{\text{phys}}$ where $\tau_{\text{phys}}$ is an experimentally determined constant. As more experiments are done and his ability to distinguish states increases (number of
relevant states increases from $2$ to $4$), perhaps he is able to postulate a higher order term proportional to $x^4$ that fits his data better. Of course, if he's going to change the Hamiltonian like this, he needs to make sure he does so in a way that doesn't leave it inconsistent with past measurements, i.e.,\ it must still have the same expectation values for $A_2$ as the thermal state for $H_0$. So if the new Hamiltonian is $H'=x^2/2\tau^2+\lambda x^4/\tau^4$ then we set $\tau$ to be the inverse (to first order in $\lambda$!) of the relation in the statistical mechanics case, $\tau=\tau_{\text{phys}}(1+6\lambda)$. Thus given his experimental data from the beginning, the agent sets his new Hamiltonian to be
\begin{equation}
H'=\frac{x^2}{2\tau_{\text{phys}}^2(1+6\lambda)^2}+\lambda \frac{x^4}{\tau_{\text{phys}}^4(1+6\lambda)^4},
\end{equation}
and he is now free to determine $\lambda$ experimentally and plug it into the above formula. 

This is in general not the end to the story, however. It could be that as he's fitting parameters that he finds $\lambda<0$ in which case the corresponding thermal state is not normalizable because the Hamiltonian is not bounded from below. Again, although we're trying to make a more detailed theory, we are still just trying to make an \textit{effective} theory so we have to keep in mind the current experimental limitations. Since the order $4$ term is at the limit of the agent's capabilities, he is free to add a yet higher term to \textit{regularize} the infinity that is wholly beyond his experimental detection. This term, called a regulator, is not thought of as physical at all, just a way to allow the state to be normalized. He may choose to add a term proportional to $x^6$. This is fine, but if he does so he has a bit more work to do because adding such a term in general changes the experimental predictions of the theory (proportional to the moments). 

So we are faced with determining how the coupling constants $\tau$ and $\lambda$ depend on $\epsilon$ in the general Hamiltonian,
\begin{equation} \label{eq:h''}
H=\frac{x^2}{2\tau^2}+\lambda \frac{x^4}{\tau^4}+\epsilon \frac{x^6}{\tau^6}.
\end{equation}
In other words, we determine how the coupling constants \textit{flow} with the regulator to keep the experimental predictions the same and to eliminate infinities. We end up with a path, $(\tau(\epsilon),\lambda(\epsilon))$ in the equivalence class. This is analogous to the regulator-dependent coupling constant flow in QFT. We can determine $\lambda(\epsilon)$ and $\tau(\epsilon)$ by the requirement that $\epsilon$ should not be detectable. This means that the expectation values of the relevant observables of the thermal state of our regularized Hamiltonian $H''$ should be independent of $\epsilon$. Without loss of generality, we define $\lambda_\text{phys}:=\lambda(0)$. Checking the least relevant observable, $A_4$, we see that $\lambda(\epsilon)$ is fixed to be $\lambda(\epsilon)=\lambda_{\text{phys}}-15\epsilon$. Checking $A_2$ we see how $\tau_\text{phys}$ is perturbed, or alternatively, plugging in $\lambda(\epsilon)$ to (\ref{eq:h''}), we can just calculate the second moment of our state to get $\tau(\epsilon)=\tau_{\text{phys}}(1+6\lambda_{\text{phys}}-45\epsilon)$. It is easy to check that our resulting Hamiltonian gives us a thermal state that flows inside the equivalence class as one varies $\epsilon$ by checking that to first order in $\lambda$ and $\epsilon$, the expectation values of the relevant observables do not depend on $\epsilon$.

This model should not be taken too seriously for obvious reasons, but it illustrates the general approach well. We will need to develop more machinery before we can move to the full quantum and quantum field setting, but before that it is instructive to extend this model to a classical field setting.

\section{Classical Fields}\label{cf}
We can easily extend the classical particle example to classical fields. By a classical field we mean a square-integrable real-valued function $\phi(x)$ over $\mathbb{R}^d$. In an informal sense we want to consider all possible such fields and define a state as a probability distribution $p(\phi)$ over all of them, extending the notion of a state as a probability distribution over $\mathbb{R}$ from the classical particle case. A Gaussian state is the natural generalization of a normal distribution,
\begin{equation} \label{eq:classicalfieldgaussian}
p(\phi) \propto e^{-\frac{1}{2}(\phi-\phi_0,A(\phi-\phi_0))},
\end{equation}
where $A$ is the covariance operator of the Gaussian, the inner product is integrating the product of the entries over $\mathbb{R}^d$, and $\phi_0$ is the center of the distribution. The covariance operator is the infinite dimensional version of a covariance matrix (used to define multi-dimensional normal distributions). As in the classical particle case, we set $\phi_0=0$ without loss of generality. Finally, as is always the case in field theory, we think of this formalism as a shorthand for an arbitrarily small, but finite, lattice model in order for normalization to be possible in general.

The extension of our noise model from the classical particle example (the stochastic map $E$) is a bit less straightforward than the extension of the definition of a state. The reason is that there are two kinds of ``fuzziness" that one should consider in the field model: field value imprecision and $\mathbb{R}^d$ distance imprecision. In the classical particle case, we only deal with distance imprecision. 

To implement both, we do the same spatial smearing as before for each field and then combine the smeared fields in another structure to implement the field value imprecision as well by integrating over all fields $\psi$ such that there is Gaussian decay for field values far from $\phi$. In symbols, 
\begin{equation} \label{eq:classicalfieldchannel}
E(p)(\phi):=\frac{1}{(2\pi h^2)^{Nd/2}}\int D\psi p(\psi)e^{-\frac{1}{2h^2}(\phi-X\psi,\phi-X\psi)},
\end{equation}
where $D\psi\equiv\prod_id\psi(x_i)$ is a shorthand for the discretized path integral and $X$ is the convolution operator properly normalized for $d$-dimensions which has integral kernel
\begin{equation}
    X(x,y)=\frac{1}{(2\pi \sigma^2)^{d/2}}e^{-\frac{1}{2\sigma^2}(x-y)^2}.
\end{equation}
Thus $\sigma$ represents the agent's precision in resolving distances and $h$ represents his precision in resolving field values.

As in the classical particle case, we imagine we are talking about a thermal state $e^{-\beta H}$ and so there is a natural association between the Hamiltonian (or the Lagrangian) and the covariance operator $A$. For the sake of general solvability, let's consider translation invariant Hamiltonians. Restricting to this subset seems reasonable in a QFT setting because we assume physics is the same everywhere in spacetime. In the statistical physics setting it may be less reasonable at least in a lattice setting because then our Hamiltonian should not be continuously translation invariant. In the translation invariant situation, $A$ and $X$ commute because they can both be diagonalized by plane waves; $A$ because the Hamiltonian is translation invariant and this condition is equivalent to being diagonalized by plane waves, and $X$ can be explicitly calculated:
\begin{equation}
X(e^{ikx})=\frac{1}{\sqrt{2\pi}\sigma}\int e^{iky}e^{-\frac{1}{2\sigma^2}(x-y)^2}dy=e^{-\frac{k^2\sigma^2}{2}}e^{ikx}.
\end{equation}
Thus we are left with a copy of the previous classical particle model for each mode labeled by wavenumber $k$. Let $a_k$ denote the eigenvalues of $A$. Looking at the form of (\ref{eq:classicalfieldgaussian}) plugged into (\ref{eq:classicalfieldchannel}) we see that for each mode, $a_k$ takes the place of $1/\tau^2$  and that the eigenvalues of $h^2X^{-2}$ take the place of $\sigma^2$.

Accordingly, the eigenvectors of $E^*\mathcal{R}^*$ are
\begin{equation}
f^m_{\textbf{k},\textbf{n}}(\phi)=\prod_{i=1}^m\frac{1}{\sqrt{n_i!}}\text{H}_{n_i}(\sqrt{a_{k_i}}\phi_{k_i})
\end{equation}
where
\begin{equation}
\phi_k:=\int \phi(x) \text{cos}(kx) dx.
\end{equation}
The label $m$ is the number of modes, \textbf{k}$=(k_1,\ldots,k_m)$ is the choice of wavenumber for each mode, and \textbf{n}$=(n_1,\ldots,n_m)$ is a choice of degree specifying the Hermite polynomial associated to each mode and thus each mode may be thought of as a harmonic oscillator. The eigenvalues are
\begin{equation} \label{eq:classicalfieldrelevance}
\eta^m_{\textbf{k},\textbf{n}}=\prod_{i=1}^m\brx{1+a_{k_i}h^2e^{k_i^2\sigma^2}}^{-n_i}.
\end{equation}
Fixing a \textbf{k} we can easily see that the relevance ratio is the same as in the classical particle case. 

From (\ref{eq:classicalfieldrelevance}) we see that there is a clear relation between the spatial precision parameter $\sigma$ and the momentum $k$. In particular, $\sigma$ essentially sets a scale for $k$; if $k$ is bigger than $1/\sigma$ then we may decide to regard this as rendering the mode irrelevant as it's raising the exponent in the denominator to a possibly large power greater than 1. Interpreting the spatial precision parameter as a momentum cutoff is one of the central insights of Wilson \cite{wilson}. For low ($k\ll 1/\sigma$) momentum modes, the spatial precision parameter is unimportant and the relevance depends only on the field value precision parameter $h$ and the degree $n$.

A simple example that we will return to in the quantum settings below is the massive classical scalar field at inverse temperature $\beta$ with Lagrangian density $\mathcal{L}=\frac{1}{2}\partial^\mu\phi\partial_\mu\phi+\frac{\mu^2}{2}\phi^2$. Assuming the fields go to zero at infinity and Fourier transforming allows us to find the eigenvalues of the corresponding covariance operator to be
\begin{equation}
a_k=\beta\sum_i \left(k_i^2+m^2\right).
\end{equation}
For $m>0$ and for $k\ll1/\sigma$ then low $n$ modes are asymptotically more relevant than higher $n$ modes as $h\to\infty$. For the massless $m=0$ case, however, all $n$ are equally relevant at $k=0$. This means that any small momentum perturbation of the hypothesis state can be detected by the agent. 

We can now determine the relevance of any general observable (to be thought of as a perturbation to the Hamiltonian) by decomposing it in terms of the eigenrelevance observables. Some perturbations to the Hamiltonian are thus more visible than others and scale differently with dimension and accuracy parameters $h$ and $\sigma$. As a brief and simple example consider the operator $B(\phi)=\int \phi(x)^2 dx$. Writing in momentum space we can decompose it as
\begin{equation}
B(\phi)=\sum_k\phi^2=\sum_k\sqrt{2}a_k^{-1}f_{k,2}^1(\phi)+\sum_ka_k^{-1}
\end{equation}
When we exponentiate we see that the second sum does not preserve the trace, i.e.,\ if we lower this term to a perturbation of the state it is proportional to the distribution (density matrix) itself and thus does not preserve the trace. We have already explicitly excluded elements in this direction from our tangent vectors so we must subtract it here in the expansion as well. Referring to the constant term as $B_0$ we calculate the relevance of the observable $B-B_0$. First note that relation (\ref{eq:relevanceratio}) when written in terms of observables $A=\Omega_\rho^{-1}(X)$ takes the form
\begin{equation} \label{eq:observerrelevanceratio}
\eta(A)=\frac{\text{Tr}(A\Omega_\rho E^\ast \mathcal{R}_\rho^\ast(A))}{\text{Tr}(A\Omega_\rho A)}.
\end{equation} 
Since we have decomposed $B$ into an expansion of eigenvectors of $E^\ast\mathcal{R}_p^\ast$ then we can just plug into (\ref{eq:observerrelevanceratio}) making use of the orthogonality of eigenrelevance observables to obtain
\begin{equation}
\eta(B-B_0)=\frac{\sum_ka_k^{-2}\eta_{k,2}^1}{\sum_ka_k^{-2}}.
\end{equation}
The relevance parameter in the numerator ensures the numerator is finite even in the continuum limit because for $k$ beyond the cutoff $1/\sigma$ the contribution gets less and less. The denominator, however, diverges and so we must make use of an ultraviolet cutoff.

The classical particle example illustrated some broad elements of renormalization and the formalism is carried further impressively with this classical field model. Going further, we would like to extend this to quantum models. But first we must take a detour to develop the language in which the following material is most comfortably cast.

\section{Phase Space Quantum Theory}\label{psqt}
Here we will briefly build some machinery which will let us naturally extend the previous examples to quantum systems. The story being briefly told here is actually a rich subject with far more details than we go into here---the goal of this Section is to allow the reader to become familiar enough with this language that we can explore the formalism already developed beyond classical models. 

\subsection{Weyl Operators}

The general idea here is to discuss one way the concept of \textit{quantization} may be formalized so that we may move away from our classical models. We consider a real vector space $V$ with a symplectic form $\sigma(f,g)$ for $f,g\in V$. A symplectic form is an antisymmetric bilinear real-valued function of two vectors. We also require $\sigma$ to be nondegenerate which means that if $\sigma(f,g)=0$ for all $f$ then $g=0$. This requirement enforces that from the perspective of the form there
is only one zero element. We define the operator $\Delta$ such that $\sigma(f,g)=(f,\Delta g)$ where $(\cdot,\cdot)$ is an inner product on our vector space. This is the classical phase space familiar from the theory of classical mechanics. The first step beyond this setting is to complexify our vector space in the canonical way such that $\sigma$ becomes a sesquilinear form (conjugate linear in the first variable).

Classical phase space is now a subset of our vector space---with $n$ position and $n$ momentum axes. A classical observable is a real linear functional on $V$. An element of the phase space can be mapped to a linear observable $\Phi_f(g):=(f,\Delta g)$ \cite{wald}. The form then defines the Poisson bracket
\begin{equation}
\{\Phi_f,\Phi_g\}=(f,\Delta g)\textbf{1},
\end{equation} 
for $f,g$ real and where $\textbf{1}(f)=1$ $ \forall f$. 

We could proceed to quantize classical observables $f \mapsto \Phi_f$, but since we want to work with bounded operators, we quantize functions $f \mapsto e^{i\Phi_f}$ instead. We define the \textit{Weyl operators} $W_f$ for all $f \in V$ which satisfy
\begin{equation}
    W_fW_g=e^{-\frac{i}{2}(f,\Delta g)}W_{f+g}\quad\text{and}\quad W_f^*=W_{-f}.
\end{equation}
When extended to $V^{\mathbb{C}}$ the relations become
\begin{equation}\label{weylops}
W_f^\ast W_g=e^{\frac{i}{2}(f,\Delta g)}W_{g-\bar{f}},\quad \text{and }\quad W_f^\ast = W_{-\bar{f}}.
\end{equation}
It is easy to see that the operators $W_f$ are unitary if $f$ is real. If this algebra can be represented by bounded operators on a Hilbert space such that the unitary groups have generators then they are $\hat{\Phi}_f$ such that
\begin{equation}\label{weylgens}
W_f=e^{i\hat{\Phi}_f},
\end{equation}
which satisfy the canonical commutation relation:
\begin{equation}
[\hat{\Phi}_f,\hat{\Phi}_g]=i(\bar{f},\Delta g)\mathbb{I}.
\end{equation}
Thus, if they exist, these are exactly the quantized version of the classical $\Phi_f$ observables. This formulation is nice not only because it makes a lot of calculations easier, but also because it demonstrates how we relatively seamlessly move from the general notion of phase space classical mechanics to the corresponding phase space quantum mechanics.

\subsection{Gaussian States and Gaussian Channels}

One large conceptual shift that we need to mention is the transition from thinking about density matrices when considering quantum mechanical systems to thinking about characteristic functions (or one of the other phase space representations of quantum mechanics). This notion works equally well in the field theoretic setting. Since the Weyl operators span the algebra of observables, a state $\rho$ is entirely specified by its action on these operators and thus by a function on phase space $f\mapsto\text{Tr}(\rho W_f)$ which defines the characteristic function. Thinking in terms of characteristic functions turns out to make our goals easier to accomplish so we will primarily do so from now on.

A \textit{Gaussian state} is one whose characteristic function is a Gaussian: $\text{Tr}(\rho W_f)=e^{-\frac{1}{2}(f-f_0,A(f-f_0))}$ for all $f\in V$. As usual, $A$ is the covariance operator, and we set $f_0=0$. This looks similar to what we defined for the classical field example, but they are actually quite different. In the classical field example we simply imagined that a state is a probability distribution over classical fields. There was no phase space and no Weyl operator. This formulation is an entirely different perspective. Thus, extended to complex numbers, a Gaussian state is one such that
\begin{equation}
\text{Tr}(\rho W_f)=e^{-\frac{1}{2}(\bar{f},A f)}
\end{equation}
for all $f\in V^\mathbb{C}$. 
This equation, the definition of the Weyl operators \eqref{weylops}, and their generators \eqref{weylgens} allows us to calculate any expectation value (or correlation function):
\begin{equation}
 \text{Tr}\left(\rho \hat\Phi_f\hat\Phi_g\right)= \frac{d^2}{dt \, ds}\text{Tr}\left(\rho W^\ast_{tf}W_{sg}\right)\biggr\rvert_{\substack{t=0 \\ s=0}}= (f,(A+\frac{i}{2}\Delta)g).
\end{equation}
This implies that the operator $A+\frac{i}{2}\Delta$ must be positive.

So we have decided to work with characteristic functions and we know what a Gaussian state is in this context. The last thing to define before we can recast the classical field material in this language and do the fully quantum examples is to define a Gaussian channel in full generality.

A general quantum channel is a CPTP map. In our classical examples they correspond to stochastic maps. Recall from the classical examples, our stochastic map was convolution with a Gaussian. This implemented a blurring to our distribution. The important feature of the convolution of a Gaussian distribution with a Gaussian was that it obtains yet another Gaussian. This motivates the definition of a Gaussian quantum channel: we want a channel that takes a Gaussian state to another Gaussian state. 

As with Gaussian states, we characterize a Gaussian channel by its action on Weyl operators:
\begin{equation}
E^\ast (W_f):=W_{Xf}e^{-\frac{1}{2}(\bar{f},Yf)}
\end{equation}
Where $X$ and $Y$ are linear operators on $V$, meaning $X^T=X^\ast$ and $Y^T=Y^\ast$. We can easily check that this map sends Gaussians to Gaussians. We use the subscript $A$ in $\rho_A$ to indicate our state is a Gaussian state with covariance operator $A$. Then looking at the characteristic function of the image of the Weyl operator under the Gaussian channel,
\begin{equation}
\begin{split}
\text{Tr}(\rho_A E^\ast(W_f)) & = \text{Tr}\brx{\rho_AW_{Xf}e^{-\frac{1}{2}(\bar{f},Yf)}}\\
& = e^{-\frac{1}{2}(\bar{f},Yf)}e^{-\frac{1}{2}(\overline{Xf},AXf)}\\
& = e^{-\frac{1}{2}(\bar{f},(X^\ast A X+Y) f)}\\
& = \text{Tr}(\rho_{X^\ast A X+Y}W_f).
\end{split}
\end{equation}
Thus, subject to the condition that $Y-\frac{i}{2}X^\ast\Delta X+\frac{i}{2}\Delta\geq 0$, we get a Gaussian state out of the channel.
\section{Classical Fields Revisited}\label{cf2}
We are now ready to prove a few general statements about the eigenrelevance observables and relevance ratios in the classical field situation. A lot of what we just developed is overkill for the classical case so there are several simplifications that happen at the outset. We will return to the quantum case, where no difficulties are alleviated, later. 

We wish to solve equation (\ref{eq:dualpicture}) for Gaussian states over arbitrarily many modes and for a Gaussian stochastic map $E$. We're going to show that we can get the eigenvalues and eigenfunctions of this relation by using a generating function. This approach is similar in spirit to the one used in the classical particle model to prove the Hermite polynomial relation we found. We will first show that the functionals
\begin{equation}
    G_f^A:=W_fe^{\frac{1}{2}(\bar{f},Af)}
\end{equation}
satisfy the relation $E^\ast\mathcal{R}^\ast_\rho(G_f^A)=G_{Hf}^A$ with $H=(1+A^{-1}X^{-1}YX^{-1})^{-1}$. We're more interested in observables than features here, so we will work with the \textit{dual metric} 
\begin{equation}
    \langle\!\langle A,B\rangle\!\rangle_\rho:=\text{Tr}(A^*\Omega_\rho (B))
\end{equation}
on observables $A$ and $B$ instead of the regular metric on features (vectors) for this Section. If $A=\Omega_\rho^{-1}(X)$ and $B=\Omega_\rho^{-1}(Y)$ for features $X$ and $Y$ then it's easy to see that this is the correct corresponding metric on observables,
\begin{equation}
    \langle\!\langle A,B\rangle\!\rangle_\rho=\text{Tr}\left(\left(\Omega_\rho^{-1}(X)\right)^*\Omega_\rho\left(\Omega_\rho^{-1}(Y)\right)\right)=\text{Tr}\left(X^*\Omega_\rho^{-1}(\Omega_\rho)\Omega_\rho^{-1}(Y)\right)=\text{Tr}(X^*\Omega_\rho^{-1}(Y))=\langle X, Y \rangle_\rho.
\end{equation}

We want to find the action of $\mathcal{R}^\ast_{\rho}$ on $G_f^A$ and then later act on the result with $E^*$. Using its definition as adjoint of $E^\ast$ and the fact that classical field operators commute, we find that for any $f$ and $g$,
\begin{equation}
\begin{split}
    \langle\!\langle \mathcal{R}^\ast_{\rho}(W_f),W_g\rangle\!\rangle_{E(\rho)} &= \langle\!\langle W_f,E^\ast(W_g)\rangle\!\rangle_{\rho}\\
    & = \langle\!\langle W_f,W_{Xg}e^{-\frac{1}{2}(g,Yg)}\rangle\!\rangle_{\rho}\\
    & = e^{-\frac{1}{2}(g,Yg)}\langle\!\langle W_f,W_{Xg}\rangle\!\rangle_{\rho}\\
    & = e^{-\frac{1}{2}(g,Yg)}\text{Tr}(\rho{W_{f+Xg}})\\
    & = e^{-\frac{1}{2}(g,Yg)}e^{-\frac{1}{2}(f+Xg,A(f+Xg))}\\
    & = e^{-\frac{1}{2}(f,Af)-(g,X^\ast Af)-\frac{1}{2}(g,(Y+X^\ast AX)g)}.
\end{split}
\end{equation}
For simplicity in what follows, define $j:=(X^\ast AX+Y)^{-1}X^\ast Af$. Now we will compute a different quantity for comparison.
\begin{equation}
\begin{split}
    \langle\!\langle W_j,W_g\rangle\!\rangle_{E(\rho)}&=\text{Tr}(\rho{E^\ast(W_{j+g})})\\
    & = \text{Tr}(\rho W_{X(j+g)})e^{-\frac{1}{2}(j+g,Y(j+g))}\\
    & = e^{-\frac{1}{2}(j+g,Y(j+g))-\frac{1}{2}(X(j+g),AX(j+g))}\\
    & = e^{-\frac{1}{2}(j,(X^\ast AX+Y)j)-\frac{1}{2}(g,(X^\ast AX+Y)g)-(j,(X^\ast AX+Y)g)}\\
    & = e^{-\frac{1}{2}(j,(X^\ast AX+Y)j)-\frac{1}{2}(g,(X^\ast AX+Y)g)-(g,X^\ast Af)}.
\end{split}
\end{equation}
Comparing we get:
\begin{equation}
    \langle\!\langle \mathcal{R}^\ast_{\rho}(W_f),W_g\rangle\!\rangle_{E(\rho)}=e^{-\frac{1}{2}(f,Af)+\frac{1}{2}(j,(X^\ast AX+Y)j)}\langle\!\langle W_j,W_g\rangle\!\rangle_{E(\rho)}.
\end{equation}
This relation is true for all $g$ so we have
\begin{equation}
    \mathcal{R}^\ast_{\rho}(W_f)=e^{-\frac{1}{2}(f,Af)+\frac{1}{2}(j,(X^\ast AX+Y)j)}W_j.
\end{equation}
Bringing the first term in the exponent to the other side we can bring it inside the argument of $\mathcal{R}_\rho^\ast$ to make $G_f^A$,
\begin{equation}
    \mathcal{R}_\rho^\ast(G_f^A)=e^{\frac{1}{2}(j,X^\ast AXj)}e^{(j,Yj)}W_j.
\end{equation}
Now we act with $E^\ast$ on both sides
\begin{equation}
    E^\ast\mathcal{R}_\rho^\ast(G_f^A)=e^{\frac{1}{2}(Xj,AXj)}W_{Xj}
\end{equation}
Now we want to take $j$ out and put it in terms of $f$ instead. Noting the identity
\begin{equation}
    \begin{split}
        Xj &= X(X^\ast AX+Y)^{-1}X^\ast Af\\
        &= X\left( (X^\ast)^{-1} (X^\ast AX+Y) \right)^{-1}Af\\
        &=X(AX+(X^\ast)^{-1}Y)^{-1}Af\\
        &=\left( (AX+(X^\ast)^{-1}Y)X^{-1} \right)^{-1}Af\\
        &=(A+(X^\ast)^{-1}YX^{-1})^{-1}Af\\
        &=\brx{A^{-1}(A+(X^\ast)^{-1}YX^{-1})}^{-1}f\\
        &=(1+A^{-1}(X^\ast)^{-1}YX^{-1})^{-1}f,
    \end{split}
\end{equation}
we define $H=\brx{1+A^{-1}(X^\ast)^{-1}YX^{-1}}^{-1}$. Thus 
\begin{equation}
    E^\ast\mathcal{R}_\rho^\ast(G_f^A)=e^{\frac{1}{2}(Hf,AHf)}W_{Hf}=G_{Hf}^A
\end{equation}
which is what we wanted to prove.

Note that $AH=H^T A$:
\begin{equation}
    \begin{split}
        AH &= \left( (1+A^{-1}(X^\ast)^{-1}YX^{-1})A^{-1} \right)^{-1}\\
        &=(A^{-1}+A^{-1}(X^\ast)^{-1}YX^{-1}A^{-1})^{-1}\\
        &=\brx{A^{-1}(1+(X^\ast)^{-1}YX^{-1}A^{-1})}^{-1}\\
        &=\brx{1+(X^\ast)^{-1}YX^{-1}A^{-1}}^{-1}A\\
        &=H^T A
    \end{split}
\end{equation}
where in the last step we used the fact that the operations of transposing and inverting commute. This means that $H$ is symmetric with respect to the scalar product $(\cdot,A\cdot)$ (which is here defined in terms of our original scalar product) because 
\begin{equation}
(x,AH(y))=(H^T A (x),y)=(AH(x),y)=(A(Hx),y)=(H(x),A(y)).
\end{equation} 
Thus with respect to this scalar product, there exists an orthonormal basis $f_k$ of $H$:
\begin{equation}
    H(f_k)=\eta_k f_k\quad\text{and}\quad(f_k,Af_l)=\delta_{kl}
\end{equation}
Finally, we obtain eigenfunctions of $E^\ast \mathcal{R}^\ast_\rho$ by taking functional derivatives of $G_f^A$ with respect to the basis functions $f_k$ and evaluating at $f=0$. We denote a functional derivative in the direction of $f_k$ by $\delta_k$ which, for a general functional $Z(f)$, is defined as
\begin{equation}
    \delta_kZ(f):=\frac{\partial}{\partial t}Z(f+tf_k)\Bigr|_{t=0}.
\end{equation}
And so finally we have
\begin{equation}
    E^\ast \mathcal{R}_\rho^\ast \left( \delta_{k_1}\ldots\delta_{k_n}G_f^A\bigr|_{f=0} \right)=\eta_{k_1}\ldots\eta_{k_n}\brx{\delta_{k_1}\ldots\delta_{k_n}G_f^A\bigr|_{f=0}} 
\end{equation}
which identifies the eigenvalues and eigenvectors in the classical field situation.

\section{Distinguishability Near Gaussian Quantum States}\label{details}
The above Section turned a messy problem into one that, although perhaps computationally difficult, is at least tractable. The quantum version of this is harder because operators do not commute, but the same sort of result is achievable (although actual calculations will remain hard).  

In what follows it is useful to define the following operator
\begin{equation}
\Theta_\rho(A)=\int_0^1\rho^s A\rho^{-s} ds.
\end{equation}
For a thermal state $\rho=\frac{1}{Z}e^{-\beta H}$, $\Theta_\rho$ gives us an alternative way to write the dual metric. Recall the definition of $\Omega_\rho$, \eqref{omegap} from Section 3, and modify it to explicitly include inverse temperature $\beta$,
\begin{equation}
    \Omega_\rho (A)=-\frac{1}{\beta Z}\frac{d}{dt}\biggr\rvert_{t=0}e^{-\beta (H+tA)}.
\end{equation}
Now, using the Dyson expansion \cite{bhatia},
\begin{equation}
    e^{A+B}-e^A=\int_0^1 e^{(1-t)A}B e^{t(A+B)}dt,
\end{equation}
we see that 
\begin{equation}
    \frac{d}{dt}\biggr\rvert_{t=0} e^{A+Bt}=\int_0^1 \frac{d}{dt}\biggr\rvert_{t=0}\left(e^{(1-s)A}Bte^{s(A+Bt)}\right)ds=\int_0^1 e^{(1-s)A} Be^{sA}ds.
\end{equation}
This relation immediately gives us
\begin{equation}
    \Omega_\rho(A)=\int_0^1\rho^sA\rho^{1-s}ds
\end{equation}
and thus
\begin{equation}
    \langle\!\langle A,B\rangle\!\rangle_\rho=\text{Tr}(A^*\Omega_\rho (B))=\text{Tr}(A^*\int_0^1\rho^s B \rho^{1-s}ds)=\text{Tr}(\rho A^*\int_0^1\rho^s B \rho^{-s}ds)=\text{Tr}(\rho A^\ast\Theta_\rho(B)).
\end{equation}
See the \hyperref[appendix]{appendix} for how $\Theta_\rho$ may be generalized to a wider class of entropies.

Just as in the classical case, we wish to solve (\ref{eq:dualpicture}) for Gaussian states. The basic plan is the same: we work with generating functions and then hopefully get a relation out of them that solves the eigensystem in one fell swoop.

We work with the generating functions
\begin{equation}
    G_f^A:=W_fe^{\frac{1}{2}(\bar{f},Af)}.
\end{equation}
Some easily calculated quantities include
\begin{equation}
    \begin{split}
        \text{Tr}\left(\rho (G_f^A)^\ast G_g^A\right)&=\text{Tr}\left(\rho e^{\frac{1}{2}(\bar{f},Af)}e^{\frac{1}{2}(g,A\bar{g})}W_f^\ast W_g\right)\\
        &=e^{\frac{1}{2}(\bar{f},Af)}e^{\frac{1}{2}(g,A\bar{g})}\text{Tr}\left(\rho e^{\frac{i}{2}(f,\Delta g)}W_{g-\bar{f}}\right)\\
        &=e^{\frac{1}{2}(\bar{f},Af)}e^{\frac{1}{2}(g,A\bar{g})}e^{\frac{i}{2}(f,\Delta g)}e^{-\frac{1}{2}(\bar{g}-f,A(g-\bar{f}))}\\
        &=e^{(f,(A+\frac{i}{2}\Delta) g)},
\end{split}
\end{equation}
where in the last step we noted that last term in the product must have a real exponent. Denoting the covariance matrix of the Gaussian state we get after applying the Gaussian channel as $B=X^\ast AX+Y$, the previous identity extends trivially to
\begin{equation}
    \text{Tr}\left(\rho E^\ast( (G_f^B)^\ast G_g^B)\right)=e^{(f,(B+\frac{i}{2}\Delta)g)}.
\end{equation}
Also,
\begin{equation}
        E^\ast(G_f^B)=E^\ast (W_f e^{\frac{1}{2}(\bar{f},Bf)})=e^{\frac{1}{2}(\bar{f},Bf)}W_{Xf}e^{-\frac{1}{2}(f,Yf)}=G_{Xf}^A.
\end{equation}

For each term in the integral and for a thermal state $\rho$, $\Theta_\rho$ conjugates a matrix with $\rho$ and $\rho^{-1}$, $X\mapsto\rho X \rho^{-1}=e^{-\beta H}Xe^{\beta H}$. This generates a group (parameterized by the variable $s$) of complex canonical transformations with the linear operator $R^A_s$ representing the group operation on phase space. It is defined by the relation
\begin{equation}
    e^{-s\beta H}W_f e^{s \beta H}=W_{R_s^A f}.
\end{equation}
where the $A$ is a reference to the covariance matrix. The defining relation also works with generating functions which we can see by first noting that $\rho$ should be invariant under the group transformation that it defines,
\begin{equation}
    \begin{split}
        \text{Tr}(\rho W_f) &= \text{Tr}(\rho W_{R^A_s f})\\
    e^{-\frac{1}{2}(\bar{f},Af)}&=e^{-\frac{1}{2}(\overline{R^A_sf},AR^A_sf)}\\
    &=e^{-\frac{1}{2}(\overline{f},(R^A_s)^TAR^A_sf)},
\end{split}
\end{equation}
and so $(R^A_s)^TAR^A_s=A$. Thus 
\begin{equation}
    e^{-s\beta H}G_f^A e^{s \beta H}=e^{-s\beta H}W_fe^{\frac{1}{2}(\bar{f},Af)} e^{s \beta H}=W_{R_s^Af}e^{\frac{1}{2}(\bar{f},Af)}=W_{R^A_s}e^{\frac{1}{2}(\overline{R^A_sf},AR^A_sf)}=G^A_{R^A_s f}.
\end{equation}
Where in the penultimate step we used the invariance of $A$. Recall that $\Theta_\rho$ involves integrating over $s$ from $0$ to $1$ so the inner product between two generating functions is
\begin{equation}
    \begin{split}
        \langle\!\langle G_f^A,G_g^A\rangle\!\rangle_\rho=\text{Tr}\left(\rho (G^A_f)^\ast \Theta_\rho(G_g^A)\right)=\int_0^1 \text{Tr}\left(\rho (G^A_f)^\ast G_{R^A_sg}\right)=\int_0^1 e^{(f,(A+\frac{i}{2}\Delta)R^A_sg)}ds.
  \end{split}
\end{equation}
So now, just as in the classical case, our goal is to compute a formula for $\mathcal{R}^\ast_\rho(G_f^A)$. Using the definition of the adjoint and several of the identities proved above we get
\begin{equation}
    \begin{split}
    \langle\!\langle \mathcal{R}^*_\rho(G^A_f),G^B_g\rangle\!\rangle_{E(\rho)}&=\langle\!\langle G_f^A,E^\ast (G_g^B)\rangle\!\rangle_\rho\\
    &=\langle\!\langle G_f^A,G^A_{Xg}\rangle\!\rangle_\rho\\
    &=\int_0^1e^{(f,(A+\frac{i}{2}\Delta)R^A_sXg)}ds.
\end{split}
\label{eq:actionofRquantum}
\end{equation}
In order to find the explicit formula for $R_\rho(G_f^A)$ we need to compare it to $\langle\!\langle G^B_h,G^B_g\rangle\!\rangle_{E(\rho)}$ as we did in the classical case. Unfortunately for our metric there may not be an analytical form as there was for the classical case. However, the form of (\ref{eq:actionofRquantum}) tells us something useful. Motivated by the classical example, taking $n$ functional derivatives of $G^A_f$ in various directions, we end up with order $n$ polynomials in the fields. What's more, if we differentiate the last equation in (\ref{eq:actionofRquantum}) a different number of times with respect to $f$ and $g$ we will get zero (when evaluated at $f=g=0$) because the exponent is linear in both $f$ and $g$. So although we may have to do this numerically in general, we know that the eigenrelevance observables and relevance ratios are generated by $G_f^A$.
\section{Quantum Particle}\label{qp}
Here we use the methods described above in our first genuinely quantum example---the quantum single particle moving in one dimension. We use $\hat{x}$ and $\hat{p}$ as position and momentum observables. A Gaussian state $\rho$ centered at the origin is one with characteristic function
\begin{equation}
    \chi_\rho(x,p)=e^{-\frac{1}{4}\text{coth}(\frac{\beta}{2})(\frac{u}{v}x^2+\frac{v}{u}p^2)}
\end{equation}
where $\beta=2\text{coth}^{-1}(uv)=\frac{1}{T}$ and $u$ and $v$ are positive parameters such that $uv\geq 1$. We've written the characteristic function in terms of $x$ and $p$, but this is just a different notation for the one used in the general formalism; in the general language we mean a phase space point is the vector $f=(x,p)$. From this, we see that our covariance matrix is 
\begin{equation}
A=\begin{pmatrix}
    \frac{1}{2}\text{coth}(\frac{\beta}{2})\frac{u}{v} & 0 \\
    0 & \frac{1}{2}\text{coth}(\frac{\beta}{2})\frac{v}{u}
\end{pmatrix}.
\end{equation}
Our Gaussian channel $Y$ is proportional to the identity on $\hat{x}$ and $\hat{p}$ but with the coefficients $\sigma_x$ or $\sigma_p$ respectively. Note the following way to represent the action of a channel: 
\begin{equation}
    \begin{split}
        E\left(e^{-H+\epsilon A}\right)&=E\left(e^{-H} + \epsilon \Omega_\rho (A)+\mathcal{O}\left(\epsilon^2\right)\right)\\
                                       &=E(e^{-H})+\epsilon E\Omega_\rho (A)+\mathcal{O}\left(\epsilon^2\right)\\
    &\approx e^{-H'+\epsilon \Omega_{E(\rho)}^{-1}E\Omega_\rho (A)}\\
    &\approx e^{-H'+\epsilon \mathcal{R}^\ast (A)}
\end{split}
\end{equation}
to first order in $\epsilon$ where $E(\rho)\propto e^{-H'}$. This suggests to us that polynomials in the observables are sent to the same order observables under $\mathcal{R}^\ast$. We find that $\hat{x}$ and $\hat{p}$ are eigenvectors for large uncertainty. As the single particle theory doesn't have any underlying spatial distances like a field theory, we only have the equivalent of field value uncertainty where we think of $\hat{x}$ and $\hat{p}$ as a field and its canonical conjugate for a single mode so for large $\sigma_x$ and $\sigma_p$ we have for $\hat{x}$
\begin{equation}
    \eta (\hat{x})=\frac{1}{1+A^{-1}Y}=\frac{1}{1+\frac{2v}{\text{coth}(\frac{\beta}{2})}\sigma_x}\approx \frac{\text{coth}(\frac{\beta}{2})}{2v}\sigma_x^{-1}\approx \frac{u}{s v}\sigma_x^{-1},
\end{equation}
where $s=\text{coth}^{-1}(uv)$, the first approximation is for large $\sigma_x$, and the second is for large temperature in case we want to take such a limit. Likewise, $\eta (\hat{p})\approx\frac{v}{s u}\sigma_p^{-1}$ in these limits.

\section{Quantum Scalar Fields}\label{qf}
Despite the difficulties of the quantum case in the general setting, if we restrict ourselves to translation invariant theories, the channel factors for each momentum mode as in the classical field example. In this way, we can use the operator $H=\brx{1+A^{-1}(X^\ast)^{-1}YX^{-1}}^{-1}$ from the general classical field treatment. 

We consider a scalar field theory with Hamiltonian 
\begin{equation}
    H=\frac{1}{2}\int dk (\Pi^2_k+\omega_k^2 \Phi^2_k)
    \label{eq:H}
\end{equation}
where $\omega_k=\sqrt{k^2+m^2}$ and $\Phi_k$ and $\Pi_k$ are, in terms of the Fourier transforms of the canonical field operators $\phi(x)$ and $\pi(x)$,
\begin{equation}
    \Phi_k=\text{Re}(\phi_k) - \frac{1}{\omega_k}\text{Im}(\pi_k) \quad \text{and} \quad \Pi_k=\text{Re}(\pi_k)+\omega_k \text{Im}(\phi_k)
\end{equation}
are real field operators that diagonalize the covariance operator associated with the massive scalar field theory. According to B\'eny and Osborne, the covariance operator has eigenvalues $\text{coth}(\beta /2)/(2\omega_k)$ and $\text{coth}(\beta /2)\omega_k/2$ for $\Phi_k$ and $\Pi_k$ respectively. The field value imprecision operator, $Y$ acts like the identity on the fields and their canonical conjugates but with different coefficients: $(Yf)(\phi,\pi)=(y_\Phi \Phi,y_\Pi \Pi)$. The distance imprecision is implemented by the same convolution channel $X$ as in the classical field setting with the convolution applied to both the fields and their canonical conjugates and is characterized by the parameter $\sigma$.

From the generating function we find that both the field operators and canonical conjugates are eigenrelevance observables. Using (\ref{eq:H}) where we can just work with the eigenvalues since our Hamiltonian is translation invariant,
\begin{equation}
   \eta (\Phi_k)= \frac{1}{1+\frac{2\omega_k}{\text{coth}(\frac{\beta\omega_k}{2})}y_\Phi e^{k^2 \sigma^2}}=\frac{\text{coth}(\frac{\beta \omega_k}{2})}{\text{coth}(\frac{\beta \omega_k}{2})+2\omega_k y_\Phi e^{k^2\sigma^2}}\approx \frac{1}{\frac{\beta\omega_k}{2}\text{coth}(\frac{\beta\omega_k}{2})+\beta \omega_k^2 y_\Phi e^{k^2 \sigma^2}}
\end{equation}
where the approximation is for large $T$. For $\Pi_k$ we have
\begin{equation}
    \eta(\Pi_k)\approx\frac{1}{\frac{\beta\omega_k}{2}\text{coth}(\frac{\beta\omega_k}{2})+\beta y_\Pi e^{k^2 \sigma^2}}.
\end{equation}

For discussion of what can be said about higher order eigenrelevance observables see the latest version of \cite{shortbeny}, Section 5 and Appendix C. There it is shown that the relevance of a polynomial of field operators is exponentially bounded by a term involving the spacial resolution $\sigma$. Thus, in an approximate sense, the expectation values of relevant observables are the $n$-point correlation functions. Also note that only the first few products of field operators and canonical conjugates have much relevance as it decreases exponentially with $n$.

\section{Renormalization}\label{renorm}
    Renormalization conditions are traditionally derived in QFT by running coupling constants as a function of the cutoff in such a way that the $n$-point correlation functions remain constant. This is exactly what we have derived by appealing to the notion of relevance, never making reference to dressed particles or to notions of which values are physical and which are bare. An agent decides which observables are relevant through the procedure we have described. With this information he learns how to modify a Hamiltonian while keeping its thermal state equivalent to the original. In the QFT setting, we found that the relevant observables are all products of field operators and their canonical conjugates up to some order and for momenta much smaller than the reciprocal of the distance precision parameter $\sigma$. Thus if a regulator is necessary, we implement the renormalization procedure by making sure our coupling constants flow in such a way that the thermal state stays in the original equivalence class.
    
    Due to the effective momentum cutoff induced by $\sigma$ noted in the classical field example, we consider field operators with mode $k>1/\sigma$ to be irrelevant. We might also consider the product of more than some number of field operators or their canonical conjugates to be irrelevant too, but this requirement is less important to the point about to be made so long as we consider at least the first several orders to be relevant. The two Hamiltonians, $H_\epsilon=\frac{1}{2}\int_{|k|<1/\epsilon}dk(\Pi_k^2+\omega_k^2\Phi_k^2)$ and $H_\sigma=\frac{1}{2}\int_{|k|<1/\sigma}dk(\Pi_k^2+\omega_k^2\Phi_k^2)$ with the regulator cutoff and the spatial precision cutoff respectively are equivalent from the perspective noted above because they differ by momentum modes bigger than $1/\sigma$. This justifies changing the bound of integration. If, however, there is an interaction term, say a $\phi^4$ interaction term, $\lambda \int dk_1\cdots dk_4 \phi_{k_1}\cdots\phi_{k_4}\delta(k_1+\ldots+k_4)$, then the parameters $m$ and $\lambda$ will have to flow with $\sigma$ so that we remain in one equivalence class (preserve the correlation functions).  As we have connected the regulator $\epsilon$ to the distance precision parameter $\sigma$ within an equivalence class, we have demonstrated that the two kinds of renormalization, statistical physics and QFT, are equivalent in this example. Informally we may think of the infinities that show up in QFT as being due to the contribution of infinitely many irrelevant features; subtracting them as we do to finitely many irrelevant features in the statistical physics situation also regularizes the theory \cite{shortbeny}.

   An increase in the precision parameter $\sigma$ and the subsequent throwing out of irrelevant terms can manifest itself in a new momentum cutoff or a change in coupling constants. However, a change in cutoff can be recast as a change in coupling constants by an appropriate rescaling of space as shown in Section E of \cite{longbeny}. In doing so, the Hamiltonian picks up an overall factor that can be compensated for by rescaling the temperature as well (which may be thought of as an imaginary time and thus spatial rescaling).
   
\section{Discussion and Outlook}\label{disc}
The papers by B\'eny and Osborne constitute a new lens through which to look at the renormalization procedure in both the statistical and QFT setting as an information theoretic process. In particular, the notion of relevance is used to identify certain directions about a thermal state as being more or less distinguishable, and the renormalization procedure is seen to be a flow of Hamiltonians whose corresponding thermal states remain indistinguishable. Two relatively nonphysical classical
examples were presented which nonetheless contained some motivating features. After reviewing phase space quantum theory, we briefly saw a proof of principle treatment of the quantum particle and the quantum scalar field theory. From the latter example we saw the condition that the renormalization flow should preserve the correlation functions is information theoretically justified. Along the way we learned some of the basics of quantum information geometry as well.

Understanding the renormalization procedure information theoretically is important because it allows us to take one step closer to identifying what is physics and what is not; it is useful in isolating where physics ends and inference begins. It is of course possible that these developments are only the beginning---that building on these ideas, a new formalism will emerge where much of what was thought to be physical phenomenon is
seen to hang together by our inferences alone.

Many future questions are listed at the end of \cite{longbeny}. In addition to this list, it may be worth investigating what utility introducing the affine connection may bring. Although hinted at in \cite{longbeny, shortbeny}, to what degree we may be able to construct nonperturbative equivalence classes with this formalism or a modification of it remains to be fully investigated.

\section*{Appendix: Operator Monotone Functions and Contractive Metrics}\label{appendix}
A function $\theta:[0,\infty)\to[0,\infty)$ is called operator monotone if, when we consider the induced function on bounded operators on our Hilbert space, $0\leq A\leq B \implies 0\leq\theta(A)\leq \theta(B)$ for all self-adjoint operators $A, B$. It turns out that there is a one-to-one correspondence between operator monotone functions $\theta$ which also satisfy $\theta(t)=t\theta(t^{-1})$ and metrics with the monotonicity property \eqref{monotone}\cite{petz}. Given such a
        $\theta$, a metric $\Omega_\rho^{-1}$ is defined via its inverse
        \begin{equation}\label{opmon}
    \Omega_\rho=\theta(L_\rho R_\rho^{-1})R_\rho
\end{equation}
where $R_\rho(A):=A\rho$ and $L_\rho(A):=\rho A$ for some matrix $A$ are superoperators that implement right and left multiplication. In the ``classical'' case, where every $\rho$ can be simultaneously diagonalized, $\theta(L_\rho R_\rho^{-1})$ becomes $\theta(\mathbb{I})$ which may be set to $1$ and \eqref{opmon} simply becomes right multiplication by $\rho$. This independence of $\theta$ in the commuting case reflects the fact that the only contractive classical information
metric is the Fisher metric. As we saw in Section 9, for the BKM metric \cite{petz,chet}, 
\begin{equation}
\theta(x)=\int_0^1 x^s ds=\frac{x-1}{\text{log}x}.
\end{equation} 
For further reading on this topic, see \cite{qig}.

\section*{Acknowledgments}
This work is a revised and improved version of my M.Sc.\ essay for the Perimeter Scholars International program submitted June 1, 2015. My advisor during the project, Ryszard Kostecki, is responsible for instigating a crisis in my world view. Largely thanks to him and his ``Quantum Fight Club'', during this year I underwent a subjectivist Bayesian transformation and radically changed the direction in which my education was headed. I thank him for unteaching me and for being an excellent r\^ole model for revolution! I would also like to thank C\'edric B\'eny for helpful discussions and correspondence regarding his work; Blake Stacey for discussions, editing, and the reference \cite{blake};
fellow PSI students [Name Redacted] and Shuyi Zhang for their help and conversations relating to physics, Linux, information security, and life; Lea Beneish for liking and hating all of the same things I do and for being my pure math counterpart. She is my biggest fan and my best friend. Finally, I thank my parents for raising me and always working to get as many opportunities for me as possible.
\bibliographystyle{cjj}
\bibliography{essay_bib}

\end{document}